\pdfoutput=1

\documentclass{aa}

\usepackage{graphicx}
\graphicspath{ {./figs/} }

\usepackage{amsmath}
\usepackage{amssymb}

\usepackage[ruled, noline]{algorithm2e}

\usepackage{booktabs}
\usepackage{txfonts}

\usepackage{natbib}

\usepackage{hyperref}
\hypersetup{pdfauthor={Alexandre Boucaud},
            pdftitle={Convolution kernels for multi-wavelength imaging}}

\let\mr=\mathrm                             
\let\bs=\boldsymbol                         

\newcommand{\x}{\bs{x}}                     
\newcommand{\y}{\bs{y}}                     
\newcommand{\n}{\bs{n}}                     
\newcommand{\h}{\bs{h}}                     
\newcommand{\Fh}{\tilde{\bs{h}}}            
\newcommand{\lap}{\bs{d}}                   
\newcommand{\Flap}{\tilde{\bs{d}}}          
\newcommand{\kab}{\bs{k}_{a,b}}             %
\newcommand{\estkab}{\hat{\bs{k}}_{a,b}}    %

\def\change#1{#1}

\def\argmin{\mathop{\mr{arg\,min}}}         
\def\norm2#1{\left\|#1\right\|^2}           

\begin{document}

\title{Convolution kernels for multi-wavelength imaging}

\subtitle{}

\author{
    A. Boucaud \inst{1, 2}
\and
    M. Bocchio\inst{1}
\and
    A. Abergel\inst{1}
\and
    F. Orieux\inst{1, 3}
\and
    H. Dole\inst{1}
\and
    M. A. Hadj-Youcef\inst{1, 3}
}

\institute{
    Institut d’Astrophysique Spatiale, CNRS, UMR 8617,
    Univ. Paris-Sud, Université Paris-Saclay, IAS, bât. 121, Univ. Paris-Sud, 91405 Orsay, France\\
    \email{alexandre.boucaud@ias.u-psud.fr}
\and
    Sorbonne Universités, UPMC Univ Paris 6 et CNRS, UMR 7095,  Institut d’Astrophysique de Paris, 98 bis bd Arago, 75014 Paris, France
\and
    Laboratoire des Signaux et Syst\`emes (Univ. Paris-Sud, CNRS, CentraleSup\'elec, Universit\'e Paris-Saclay), 91192, Gif-sur-Yvette, France.
}

\date{Received June 9, 2016; accepted September 5, 2016}

\abstract{
Astrophysical images issued from different instruments and/or spectral bands
often require to be processed together, either for fitting or comparison
purposes. However each image is affected by an instrumental response, also
known as PSF, that depends on the characteristics of the instrument as well as
the wavelength and the observing strategy. Given the knowledge of the PSF in
each band, a straightforward way of processing images is to homogenise them all
to a target PSF using convolution kernels, so that they appear as if they had
been acquired by the same instrument. We propose an algorithm that generates
such PSF-matching kernels, based on Wiener filtering with a tunable
regularisation parameter. This method ensures all anisotropic features in the
PSFs to be taken into account. We compare our method to existing procedures
using measured \textit{Herschel}/PACS and SPIRE PSFs and simulated
\textit{JWST}/MIRI PSFs. Significant gains up to two orders of magnitude are
obtained with respect to the use of kernels computed assuming Gaussian or
circularised PSFs. A software to compute these kernels is available at
\url{https://github.com/aboucaud/pypher}.
}

\keywords{
    point-spread function --
    PSF homogenisation --
    convolution kernels
}

\maketitle


\section{Introduction}
\label{sec:intro}

The point-spread function (PSF), also known as \textsl{beam}, is one of the main
characteristics of any astronomical imager. It is a model of the diffraction
pattern resulting from the interaction between the electromagnetic radiation and
the instrument optics and detectors at every wavelength. Since most instruments
operate on a single or a series of bandpasses (through e.g. filters), the
resulting effective PSF is an integral of the monochromatic PSFs over the
wavelength range, weighted by the instrumental throughput and the source energy
distribution of a given astronomical object. A more accurate model can
even include convolutional effects such as guiding errors, trailing effects from
a scanning mode, smearing by the detector response or even non-convolutional
effects like the brighter-fatter effect. Once imaged, these model PSFs exhibit a
complex shape, including anisotropy, wings and spikes that extend far from the
center.
Another classic feature of the PSF derived from the laws of optics is the
radially oscillating pattern of the response, (especially in the monochromatic
case) creating a series of peaks and valleys. These secondary peaks can account
for a non negligible amount of the total power of the PSF. For ground-based
astronomy though, the atmospheric turbulence creates a smearing that
redistributes the power of these peaks and valleys and enables the PSF to be
modeled by simple analytic profiles like 2D Gaussians. On the contrary, space
telescopes can benefit \change{from} a much higher resolution at the expense of a full
complexity of the PSF.
To cite a few examples, the effective PSF of \textit{IRAS} maps was elliptical
due to the scanning strategy, so the angular resolution was strongly anisotropic,
with ratios up to 1:6
\citep[\textit{e.g.} 0.75' $\times$ 4.6' at 25\,$\mu$m, from][]{wheelock1994iras}.
More recently, the effective PSFs of the
\textit{Planck}/HFI\footnote{\url{http://planck.esac.esa.int}} maps appeared to
have an ellipticity in a range 1.04 to 1.4, depending on the spectral band
\citep{ade2011planck}, and the PSF of the
PACS\footnote{\url{http://herschel.esac.esa.int}} photometer
\citep{2010A&A...518L...2P} onboard the \textit{Herschel} satellite, characterised by
\citet{lutz2012pacs}, showed a narrow core, a tri-lobe pattern and Knotty
structured diffraction rings.
As we push both optical performances and detector capabilities of the future
missions towards the boundaries, the optical design highly increases in
complexity. Hence, for upcoming space surveys
(Euclid\footnote{\url{http://www.euclid-ec.org/}},
WFIRST\footnote{\url{http://wfirst.gsfc.nasa.gov/}}) or
observatories (Athena\footnote{\url{http://www.the-athena-x-ray-observatory.eu/}}
or JWST\footnote{\url{http://http://www.jwst.nasa.gov/}}),
the characterisation and processing of elaborated PSFs become a crucial task.\\

Most of astrophysical studies necessitate multi-wavelength observations, either from
multiple bands/filters within an instrument or from various instruments and
telescopes.
The different maps are affected by a particular PSF and the
pixel-based data comparison cannot be straightforward. However, a technique
widely used in multi-band photometry is to perform the measurements on PSF
homogenised data, that is to select a dataset as reference (usually the one with
the worst resolution, or wider PSF) and transform the other datasets so they are
PSF-matched with the reference PSF, a technique called PSF homogenisation or
PSF matching \citep[see e.g.][]{bertin2002terapix,gordon2008,darnell2009dark,
hildebrandt2012cfhtlens}.
PSF homogenisation is usually achieved by convolving the image with a kernel
that is generated from the PSF corresponding to the image and the reference PSF.
In the literature, one can distinguish between parametric kernels which use a
fit of an analytic model to each PSF (Moffat, multiple Gaussians, etc.) or their
decomposition on a proper basis (e.g. Gauss-Hermite polynomials, or shapelets),
and results in an analytic expression for the kernel
\citep[e.g.][]{kuijken2008gaap,hildebrandt2012cfhtlens} ; and non parametric
methods that use pixel information from the image \citep[e.g.][]{alard2000} or
adopt effective PSF images \citep[e.g.][]{gordon2008,aniano2011} to compute the
kernels.\\

With the purpose of taking into account the full complexity and angular
extension of space instruments' PSFs, we
address the creation of PSF-matching kernels for multi-wavelength studies.
We then present two usecases for these kernels, one based on the
\textit{Herschel} satellite data, and a second on simulations for the MIRI
instrument of \textit{JWST}. We also deliver a program called \texttt{pypher}
that computes the kernels given two PSF images (see Appendix~\ref{sub:pypher}).
This code has initially been developed in preparation for the Euclid mission
\citep{laureijs2010euclid}.

In Section~\ref{sec:method}, we describe the algorithm for the generation of
convolution kernels used to match the resolution of images.
In Section~\ref{sec:herschel}, we assess the improvement brought by
these kernels on the multi-wavelength study of dust properties with the
\textit{Herschel} satellite, and show in Section~\ref{sec:jwst} the
reconstruction power of such kernels on PSF simulations of the future
\textit{JWST} satellite, before summarising this work in
Section~\ref{sec:conclusion}.


\section{Kernel generation}
\label{sec:method}

\subsection{Data Model}
\label{sub:data_model}

Let's first consider an astrophysical image $\y$, observed with an instrument
modeled as a linear invariant system:
\begin{equation}
    \y = \h \ast \x + \n\,,
\end{equation}
where $\h$ is the PSF convolved with the unknown sky $\x$, $\n$ is the image
noise and $\ast$ stands for the discrete convolution \change{\citep[see e.g.][]{gonzalez2008digital}}.

Given two PSF models $\h_a$ and $\h_b$, where $a$ and $b$ refer to different
frequency bands from the same or various instruments, the
process we are interested in, referred to as PSF-matching, is to
transform the image $\y_a$ acquired at the angular resolution of $\h_a$
\begin{equation}
    \y_a = \h_a \ast \x + \n_a\,,
\end{equation}
into an image $\y_{a,b}$ at the angular resolution of $\h_b$
\begin{align}
    \y_{a,b}
        &= \kab \ast \y_a\,,
    \label{eq:matching}\\
        &\simeq \h_b \ast \x
\end{align}
where $\kab$ is the matching kernel from $\h_a$ to $\h_b$.\\

This paper presents a linear algorithm that computes the kernel $\kab$ to
produce the image $\y_{a,b}$ from the original image $\y_a$ through a
convolution.
To this end, we need to construct the kernel $\kab$ such that
\begin{equation}
    \h_b = \h_a \ast \kab \,.
    \label{eq:linsys}
\end{equation}

\subsection{Kernel generation}
\label{sub:kernel_generation}

For such linear systems as equation~\eqref{eq:linsys}, one can seek an estimate
of $\kab$, denoted $\estkab$, that minimises the least squares criterion $J$
\begin{align}
    \estkab
        &= \argmin_{\kab} J(\kab)\,,
            \label{eq:jacob}\\
        &= \argmin_{\kab} \norm2{\h_b - \h_a \ast \kab}\,.
            \label{eq:estimk}
\end{align}
However, the presence of the convolution makes the system ill-posed for the
inversion, hence the solution to equation~\eqref{eq:estimk} is not stable.
The only way to stabilise the solution is to add information. For the considered
system, we use a technique called \textit{regularisation}. We choose a
$\ell_2$ norm in order to have a linear estimator \change{and use Fourier filtering}, and
penalise the high frequencies where we expect the noise to dominate, using a high-pass filter
$\lap$. This corresponds to adding a relative degree of smoothness between
values of neighboring pixels.
\begin{equation}
    J(\kab) = \norm2{\h_b - \h_a \ast \kab} + \mu\norm2{\lap \ast \kab}\,,
    \label{eq:jreg}
\end{equation}
where $\lap$ is the second-order differential operator (\textit{i.e.} two-dimensional
Laplacian matrix)
\begin{equation}
    \lap =
        \begin{bmatrix}
           ~0 & -1 & ~0\\
           -1 & ~4 & -1\\
           ~0 & -1 & ~0
        \end{bmatrix}
\end{equation}
and $\mu$ the regularisation parameter, which tunes the balance between the data
fidelity and the penalisation. \change{Other norms, such as $\ell_1$,
$\ell_2\ell_1$ or $TV$ (Total Variations) are known to better preserve the image
details, but produce non linear estimators that require iterative algorithms to
solve.}\\

Denoting the Fourier transform of any two-dimensional vector $\bs{u}$ by
$\tilde{\bs{u}}$, the convolution theorem states that the real space convolution
is equivalent to a termwise product in Fourier space
\begin{equation}
    \h \ast \bs{k} \Leftrightarrow \Fh \odot \tilde{\bs{k}}\,,
\end{equation}
where $\odot$ symbolises the termwise product between vectors/matrices.

Under these assumptions, the cancellation of the first gradient of the criterion
\eqref{eq:jreg} leads to the classical regularised mean square solution of
equation \eqref{eq:estimk} in Fourier space
\begin{equation}
    \tilde{\hat{\bs{k}}}_{a,b} = \bs{w} \odot \Fh_b
    \label{eq:fresult}
\end{equation}
where $\bs{w}$ is a Wiener filter with high-frequency penalisation
\begin{equation}
    \bs{w}(\mu) = \frac{\Fh_a^{\dag}}
                  {|\Fh_a|^2 + \mu|\Flap|^2}
             \quad;\quad \mu \neq 0
    \label{eq:wiener}
\end{equation}
and $\bs{\Fh_a}^{\dag}$ stands for the complex conjugate of matrix $\bs{\Fh_a}$.\\

The real-space convolution kernel $\kab$,
is eventually obtained via the inverse Fourier transform of
equation~\eqref{eq:fresult}. For two instruments $a$ and $b$, this kernel is
thus only parametrised by the regularisation parameter $\mu$. The optimal
balance between the data and the penalisation is found by setting $\mu$ to the
inverse of the signal-to-noise ratio (SNR) of the homogenised image, $\y_a$ in
this case.

We provide with this paper the \texttt{pypher} program (see
Appendix~\ref{sub:pypher}), an implementation of the Algorithm~\ref{algorithm}
presented below, that computes $\kab$.

\begin{algorithm}
\DontPrintSemicolon
    \SetKwInOut{Input}{inputs}\SetKwInOut{Output}{output}
    \Input{$\h_a$ 2D array of size N$_a$ $\times$ N$_a$ and pixel scale p$_a$,
           \newline $\h_b$ 2D array of size N$_b$ $\times$ N$_b$ and pixel scale p$_b$,
           \newline angles $\alpha_a$ and $\alpha_b$ (see Appendix~\ref{sub:pypher})
           \newline regularisation factor $\mu$.}
    \Output{$\kab$ 2D array of size N $\times$ N and pixel scale p.}
    \BlankLine

    \tcc*[r]{PSF warping:}
    N = N$_b$ ; p = p$_b$\;
    \BlankLine
    \For{$i$ in \{$a$, $b$\}}{
        \If{$\alpha_i \neq 0$}{
            rotate $\h_i$ through an angle $\alpha_i$\;
        }
    }
    \BlankLine
    \If{p$_a \neq$ p}{
        rescale $\h_a$ to the pixel scale p\;
    }
    \BlankLine
    \For{$\bs{u}$ in \{$\h_a$, $\lap$\}}{
        \eIf{size($\bs{u}$) < N $\times$ N}{
            pad $\bs{u}$ with zeros to a size of N $\times$ N\;
        }{
            trim $\bs{u}$ to a size of N $\times$ N\;
        }
    }
    \BlankLine
    \tcc*[r]{Wiener filter:}
    \For{$\bs{u}$ in \{$\h_a$, $\lap$\}}{
        compute the OTF\footnotemark[7] of $\bs{u}$: $\tilde{\bs{u}}$\;
    }

    compute $\bs{w}(\mu)$ following equation~\eqref{eq:wiener}\;
    \BlankLine
    \tcc*[r]{Kernel:}
    compute the discrete Fourier transform of $\h_b$: $\Fh_b$\;
    compute $\tilde{\hat{\bs{k}}}_{a,b}$ following equation~\eqref{eq:fresult} via a termwise product\;
    inverse Fourier transform $\tilde{\hat{\bs{k}}}_{a,b}$ to obtain the kernel $\kab$\;

    \caption{Matching kernel generation recipe}
    \label{algorithm}
\end{algorithm}

\footnotetext[7]{The optical transfer function (OTF) is the discrete Fourier
transform of a signal which has been translated so that its peak value is the
first vector entry (\textit{i.e.} Im[0, 0] for an image).}
\addtocounter{footnote}{+1}


\section{Impact on \textit{Herschel} data analysis}
\label{sec:herschel}

\defcitealias{2011PASP..123.1218A}{ADGS11}

The \texttt{pypher} kernels allow us to convolve multiple astronomical images to
a common angular resolution. In order to preserve all the information during
this process, a good knowledge of the effective PSFs of the instruments used is
required. We focus here on the particular analysis of \textit{Herschel}
photometric images which have been widely used in the last few years to measure
the dust temperature and the spectral index $\beta$ across many astronomical
objects from multi-band imaging with PACS and SPIRE.

Given the uncertainties on the PSFs of PACS and SPIRE instruments, since the
beginning of the operational period of \textit{Herschel} and for a few years,
convolutions from PACS to SPIRE angular resolution have been performed assuming
Gaussian PSFs with a given FWHM estimated from dedicated observations of
asteroids. In 2011, a better characterisation of the instruments' PSFs allowed
\citet[ADGS11]{2011PASP..123.1218A} to develop a method to construct convolution
kernels assuming circular PSFs. While a full analysis of the PSF of the SPIRE
instrument has been finalised recently by the Instrument Control Centre
\citep[ICC,][]{SPIRE_PSF}, a final PSF characterisation for the PACS instrument
has not been released yet by the ICC \citep{PACS_PSF}. Parallel work has been
performed by \citet{paper_PSFs}, who have computed PACS effective
PSFs\footnote{these PSFs are publicly available at\\
\url{http://idoc-herschel.ias.u-psud.fr/sitools/client-user/Herschel/project-index.html}.}
from the combination of Vesta and Mars dedicated observations, which will be
used hereafter. The current knowledge of both PACS and SPIRE PSFs, and the
\texttt{pypher} code allow us to construct effective convolution kernels with an
unprecedented precision. In this section we show how and to what extent the use
of different convolution kernels can affect the results.

\subsection{Herschel PSFs and kernels}
\label{sec:Herschel_PSFs}

For each \textit{Herschel} band, $\lambda$, we define four PSF images:
\begin{enumerate}
\item the effective PSF ($E_{\lambda}$), taken from \citet{paper_PSFs} for PACS
and \citet{SPIRE_PSF} for SPIRE ;
\item the Gaussian PSF ($G_{\lambda}$), computed fitting a 2D Gaussian profile
to $E_{\lambda}$. $G_{\lambda}$ has the same FWHM as $E_{\lambda}$ but does not
account for secondary lobes and faint structures ;
\item the circular PSF ($C_{\lambda}$), computed from $E_{\lambda}$ by averaging
the image intensity in annular bins.
All information on the asymmetry of the PSF is then lost ;
\item the Aniano PSF ($A_{\lambda}$), a circular PSF from the
\citetalias{2011PASP..123.1218A} paper.
\end{enumerate}
For each of the first three types of PSF, we define the corresponding matching
kernels between the bands $\lambda_1$and $\lambda_2$ as
\begin{equation}
    K^X_{\lambda_1, \lambda_2} \quad \text{with} \quad X \in \{E, G, C\}
    \label{eq:kernels}
\end{equation}
and compute them with \texttt{pypher}. We also consider the Aniano kernels
defined here as $K^A_{\lambda_1, \lambda_2}$ and computed in
\citetalias{2011PASP..123.1218A} via a different method than that used in
this work.

The homogenised images can therefore be denoted by
\begin{equation}
    X_{70, 350} = E_{70} \ast K_{70, 350}^{X}\,.
\end{equation}

\begin{figure}[t]
    \centering
    \includegraphics[width=0.45\textwidth]{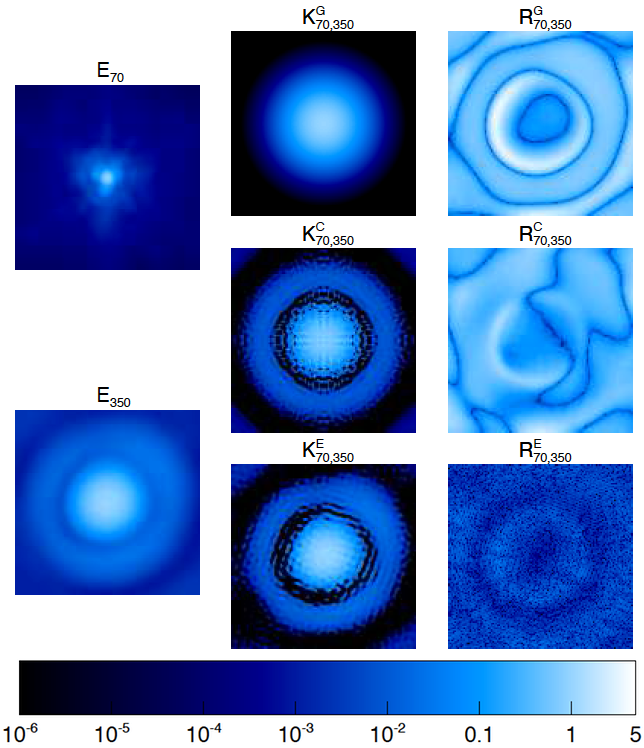}
    \caption{Effective PACS 70 $\mu$m and SPIRE 350 $\mu$m
    PSFs on the left. The central column shows Gaussian, circular and
    effective kernels for these PSFs computed following the procedure described
    in Section~\ref{sec:Herschel_PSFs}. The corresponding relative residual
    images (see equation~\ref{eq:residuals}) are displayed in the right column.
    The bottom kernel and residual image correspond to this work.
    All images are $120'' \times 120''$.}
    \label{fig:PSF_kernels}
\end{figure}

\subsection{Kernel comparison}
\label{sub:kern_comparison}

In this paragraph we assess the impact of using approximations of effective PSFs
in the homogenisation process, which is directly related to the creation of the
convolution kernel. The effective kernel $K^E_{\lambda_1, \lambda_2}$ being the
target of our kernel generation algorithm, it is expected to produce better
results than the other types described above.

To compare the different types of kernels, we chose to measure the difference
between the effective PSF at $\lambda_1$ matched to $\lambda_2$ and the
effective PSF at $\lambda_2$.
We define the relative residuals $R^X$ as
\begin{equation}
    R^X_{\lambda_1, \lambda_2} =
        \frac{
            |E_{\lambda_2} - E_{\lambda_1} \ast K^X_{\lambda_1, \lambda_2}|
        }{
            E_{\lambda_2}
        }.
    \label{eq:residuals}
\end{equation}
for each type of kernel.\\

For this comparison test, we consider the matching of the PACS 70 $\mu$m PSF
to the resolution of SPIRE 350 $\mu$m. In the remainder of this paragraph, we
refer to these bands as 70 and 350.

The two images on the left of Figure~\ref{fig:PSF_kernels} represent the
effective PSFs of PACS 70 $\mu$m: $E_{70}$ and SPIRE 350 $\mu$m: $E_{350}$. The
central column shows the kernels computed with \texttt{pypher} from Gaussian
(top) and circular (middle) approximations of the effective PSFs, as described
in Section~\ref{sec:Herschel_PSFs}, and directly from the left-side PSFs
(bottom). On these kernel images, one can see the characteristics of the input
PSFs: the Gaussian approximation has a single lobe, the circular one is
axisymmetrical and presents a second lobe, and the last one has two lobes and
the general shape of $E_{350}$. Next to these kernels, the associated
homogenisation residual images are displayed (see equation \ref{eq:residuals}
for computation).

\begin{figure}[t]
    \centering
    \includegraphics[width=0.45\textwidth]{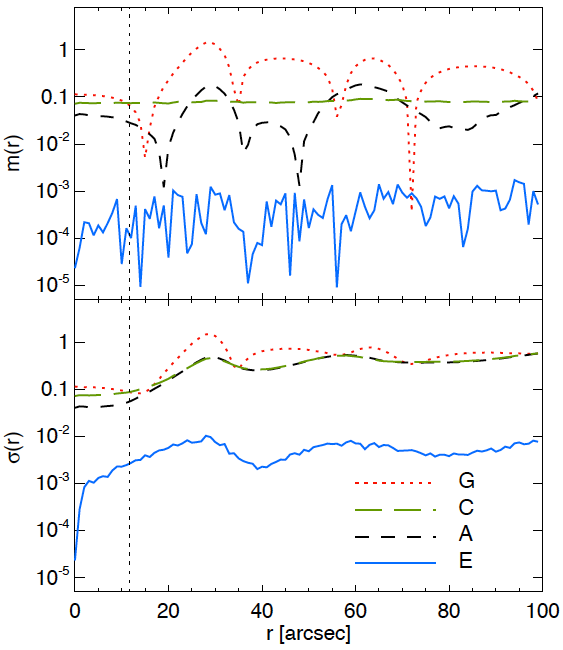}
    \caption{First ($m(r)$, top panel) and second ($\sigma(r)$, bottom panel)
    polar moments of the residuals (right column images of
    Figure~\ref{fig:PSF_kernels} + $R^A_{70, 350}$) as a function of the distance
    $r$ to the image center. The different lines represent the kernel types used
    for the homogenisation: Gaussian (red dotted), circular (green long dashed),
    \citetalias{2011PASP..123.1218A} (black dashed) and effective (blue solid,
    this work). The vertical black dotted line indicates the HWHM of the SPIRE
    350 $\mu$m PSF.}
    \label{fig:momResTot70_350}
\end{figure}

Both $R^G_{70, 350}$ and $R^C_{70, 350}$ have residuals on the
order of 10\% within the first lobe of $E_{350}$ (central region of radius equal
to the HWHM $\simeq 12''$). Outside of that region, the reconstruction from
these two kernels is even worse. In particular, the extinction ring that marks
the transition between the first and second lobe does not exist in the Gaussian
case and is slightly shifted due to azimuthal averaging in the circular case,
which leads to a big residual error in both cases (white circle on
$R^G_{70, 350}$ and $R^C_{70, 350}$). Both kernel and residual images from
\citetalias{2011PASP..123.1218A}, which are not shown on
Figure~\ref{fig:PSF_kernels}, present very similar behaviour to the circular
approximation. Using the kernel constructed with effective PSFs,
$R^E_{70, 350}$, we obtain very homogeneous residuals of the order of 0.1\%.

To analyse in more details these residuals, we introduce the first $m$ and
second $\sigma$ polar moments of the residuals:
\begin{align}
    m(r) &=
        \frac{1}{n_{\rm ang}} \sum_{i=1}^{n_{\rm ang}}
            R^X(r, \theta_i),\nonumber\\
    \sigma(r) &=
        \sqrt{\frac{1}{n_{\rm ang}} \sum_{i=1}^{n_{\rm ang}}
            \left[ R^X(r, \theta_i) \right]^2}\,,
\end{align}
where $R^X(r, \theta_i)$ is the intensity of the residual image
(computed using the kernel $X$) at a distance $r$ from the image centre and at
an angle $\theta_i = 2\pi i /n_{\rm ang}$, with $n_{\rm ang} = 100$.
The moments express the intensity and dispersion of the residuals along
the PSF radius.

Figure~\ref{fig:momResTot70_350} shows these two values computed on the residual
images for the four kernel types. As previously stated, the Gaussian case
(dotted red lines) is only stable within a circle of radius equal to the HWHM of
$E_{350}$. At further distance it shows very high first and second moments,
close to unity, and establishes a very poor matching.
The circular case (long-dashed green) exhibits a constant first moment
below 10\% at all distances, better than the Gaussian case. This is mainly due
to the computation of the first moment that is very similar to the circularising
process and averages out the measurements.
However, the asymmetry and local structures of $E_{70}$ and $E_{350}$ are lost
and the second moment is comparable to that of $R^G_{70, 350}$.
\citetalias{2011PASP..123.1218A} used narrower versions of PACS and SPIRE PSFs
than those used in this work to produce the kernels. Both first and second
moments (dashed black) therefore present bumps at the position of the lobes.
Finally, the effective case (blue) best matches $E_{350}$. In amplitude, the
first moment is $\lesssim 0.1\%$ and the second moment $\lesssim 1\%$ at all
distances, even if we note that the second moment presents the two bumps
observed earlier. By comparison, these moments are two orders of magnitude lower
than that of the other three cases.

This concludes in a major improvement in
using \texttt{pypher} kernels with effective PSFs with respect to Gaussian,
circular or \citetalias{2011PASP..123.1218A} kernels from an image processing
standpoint. Next we test again these kernels on the determination of meaningful
parameters from data or simulations.

\subsection{Dust properties study}

\textit{Herschel} observations are often used to retrieve information on dust
properties in our Galaxy as well as in local galaxies. In this section we show
how the choice of PSFs to construct the convolution kernels can affect
pixel-by-pixel measurements.

\begin{figure}[t]
    \centering
    \includegraphics[width=0.45\textwidth]{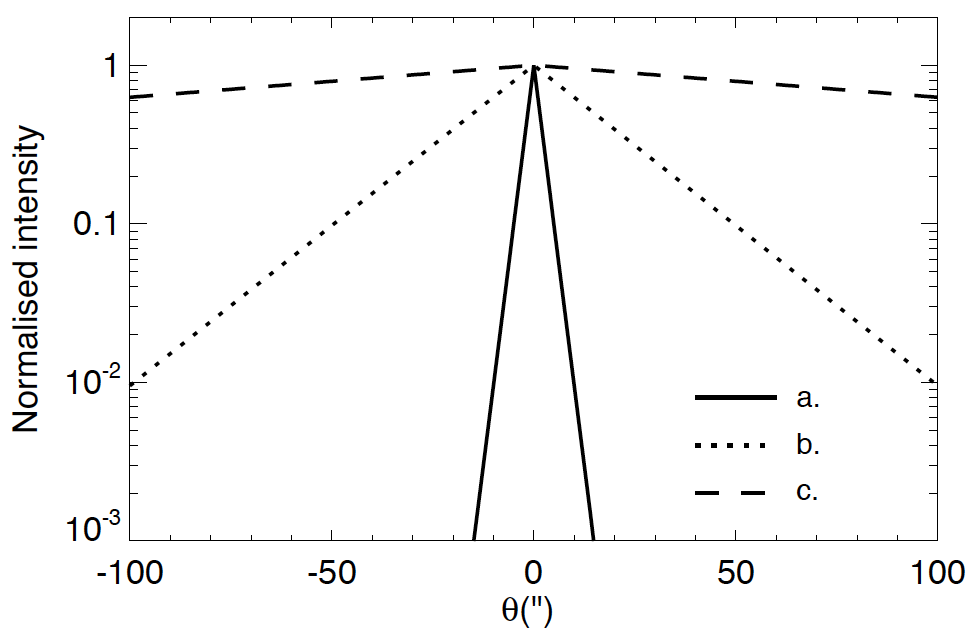}
    \caption{Intrinsic dust profiles (see equation \ref{eq:dustprofil}) for scenarios
    \textit{a.}, \textit{b.} and \textit{c.} listed in
    Table~\ref{table:scenarios}, corresponding, respectively, to characteristic
    scale heights of $z_d = 0.1$, 1 and 10 kpc. These profiles are then
    convolved with \textit{Herschel} PSFs to simulate an edge-on galaxy observed
    in different bands.}
    \label{fig:profiles}
\end{figure}

We consider an edge-on galaxy at $D \sim 10$\,Mpc from us with an intrinsic
vertical profile given by:
\begin{equation}
    I_{\rm d}(z) = I_{\rm d}(0) \exp\left(\frac{-z}{z_{\rm d}}\right),
    \label{eq:dustprofil}
\end{equation}
where $z_{\rm d}$ is the scale height of the vertical dust distribution. We
introduce the angular distance $\theta \simeq z/D$ and its characteristic value
$\theta_{\rm d} \simeq z_{\rm d}/D$. Three scale heights are examined, $z_{\rm
d} = $ 0.1, 1 and 10 kpc (scenarios \textit{a.}, \textit{b.} and \textit{c.} from
Table~\ref{table:scenarios}), corresponding to $\theta_{\rm d} \simeq 2, 20$ and
$200''$. Their intrinsic dust profile is shown on Figure~\ref{fig:profiles}.

We convolve the intrinsic dust profile to the effective PSFs of PACS 70, 100 and
160 $\mu$m and SPIRE 250 and 350 $\mu$m, while keeping the pixel size at
$1''$. We then make the basic assumption that the dust in the whole
galaxy has a temperature of $T_{\rm d} = 20$\,K and spectral index
$\beta = 1.6$ (following \citealt{2014A&A...571A..11P}), and rescale the
convolved models accordingly.

In order to simulate real data, we add Gaussian statistical noise to the models
and consider three different values of signal-to-noise ratio (SNR), $10^{4}$,
$10^{2}$, $10$ (scenarios \textit{b.}, \textit{b.1} and \textit{b.2} from
Table~\ref{table:scenarios}), with respect to the dust emission at the peak
position ($z = 0$). These images are then homogenised to the resolution of SPIRE
350 $\mu$m using the four kernel types described in
Section~\ref{sec:Herschel_PSFs} and resampled with a common pixel size of
$10''$.

Finally, using a minimum $\chi^2$ method, we fit the multi-wavelength data on a
pixel-by-pixel basis to a modified blackbody:
\begin{equation}
    I_{\nu} = \tau_{\nu_0} (\nu / \nu_0)^{\beta} B_{\nu}(T),
\end{equation}
where $\tau_{\nu_0}$ is the optical depth at the reference frequency $\nu_0$
and $B_{\nu}(T)$ is the blackbody radiation for a grain at temperature $T$.
\begin{figure}[t]
    \centering
    \includegraphics[width=0.45\textwidth]{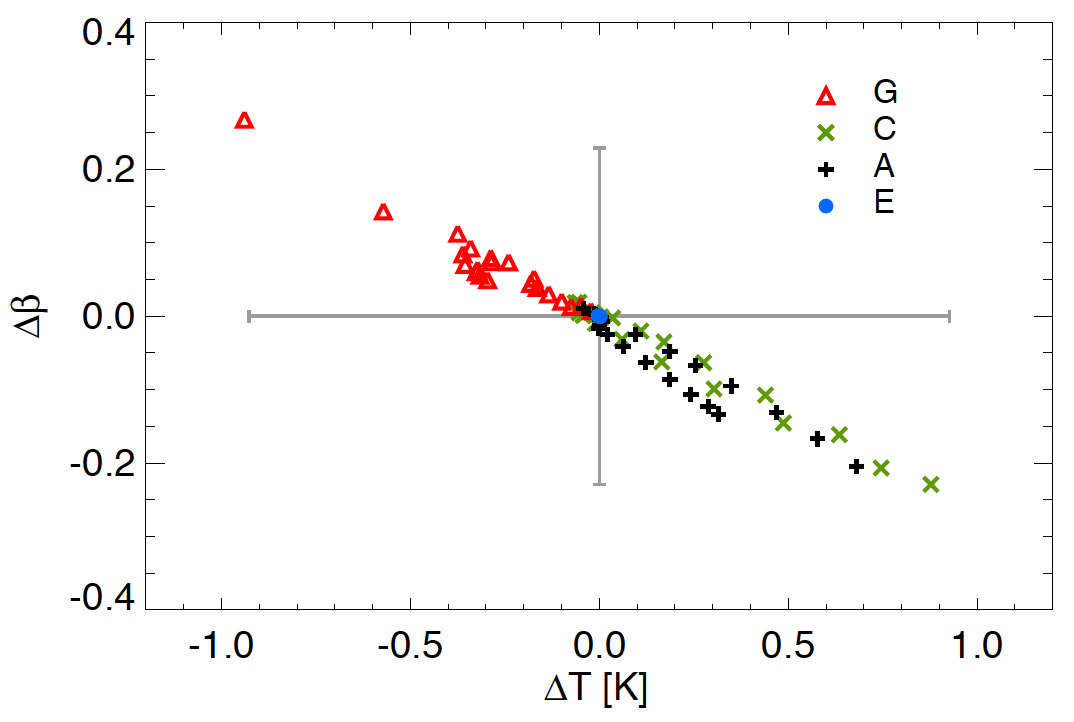}
    \caption{$\Delta\beta$ parameter as a function of $\Delta T$ for the four
    kernel types, using an edge-on galaxy with the dust profile \textit{b.}
    ($z_d = 1$ kpc, SNR = $10^{4}$). For a given
    kernel type, there are 21 data points, each of them corresponding to a measurement on a single pixel along the galaxy profile.
    The spread thus represents the systematic error on the $\beta - T$ measurement
    for this kernel type. Error bars on both axis at the centre indicate the
    statistical errors on $\beta$ and $T$ obtained from the $\chi^2$-fitting
    routine.}
    \label{fig:beta_temp}
\end{figure}

We define $\Delta\beta$ and $\Delta T$ the deviations from the reference dust
spectral index $\beta$ and temperature $T_{\rm d}$, respectively.
Figure~\ref{fig:beta_temp} shows
$\Delta\beta$ as a function of $\Delta T$ for scenario \textit{b.} where $z_{\rm
d} = 1$\,kpc and SNR = $10^{4}$, for the different kernel types. Except for the
case where effective PSFs are used, systematic discrepancies are present (up to
$\Delta T \pm 1$\,K and $\Delta \beta \pm 0.3$), comparable in amplitude to the
statistical errors (horizontal and vertical bars on Figure~\ref{fig:beta_temp}),
and a spurious strong negative correlation appears between dust temperature and
$\beta$ parameter.

\begin{table}[b]
\caption{Scale height, $z_{\rm d}$, and signal-to-noise ratio (SNR) adopted for
the simulation of dust profiles of an edge-on galaxy.
\vspace{.2cm}}
\label{table:scenarios}
\centering
\begin{tabular}{ccc}
    \toprule
    Scenario & $z_{\rm d}$ (kpc) & SNR \vspace{.05cm}\\
    \midrule
    $a.~$ & ~0.1 & $10^{4}$\\
    $b.~$ & ~1.0 & $10^{4}$ \\
    $c.~$ & 10.0 & $10^{4}$ \\
    $b.1$ & ~1.0 & $10^{2}$ \\
    $b.2$ & ~1.0 & $10^{1}$ \\
    \bottomrule
\end{tabular}
\end{table}

\begin{figure}[t]
    \centering
    \includegraphics[width=0.5\textwidth]{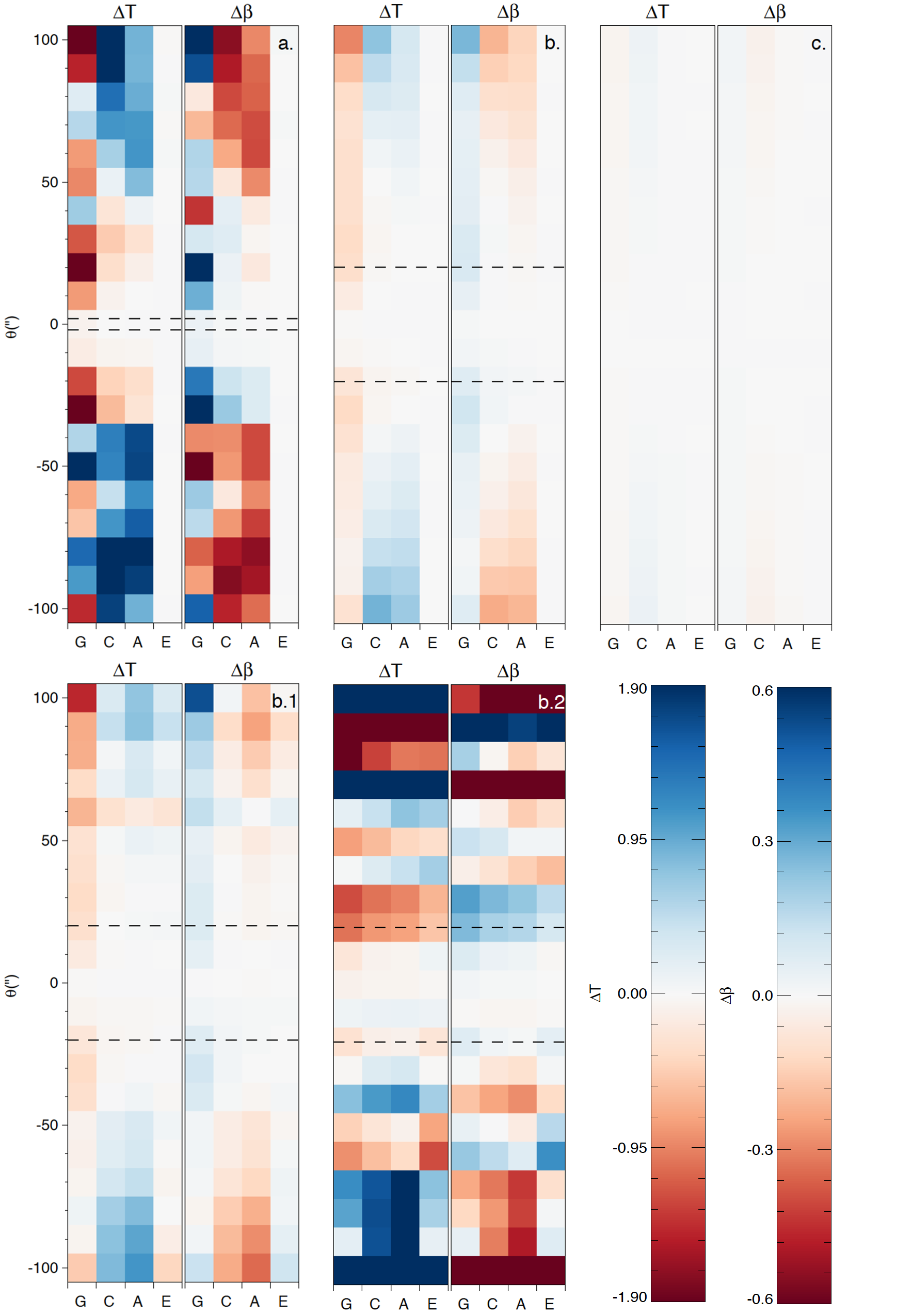}
    \caption{
    Heat maps of $\Delta T$ and $\Delta \beta$ measured on a simulated edge-on
    galaxy at various angular distances $\theta$ from the center, as a function
    of the kernel type used. Each panel represents a scenario from
    Table~\ref{table:scenarios}. Dashed lines indicate the scale height of the
    dust profile $z_{\rm d}$. This figure highlights the very small scatter in
    both $\Delta T$ and $\Delta \beta$ when using the kernels with effective
    PSFs (column E on each plot), until the SNR becomes to low as in panel $b.2$.}
    \label{fig:mosaic_deltaT}
\end{figure}

In order to show where this effect is most significant within the galaxy profile, we
illustrate on Figure~\ref{fig:mosaic_deltaT} the quantities $\Delta T$ and $\Delta
\beta$ at various angular distances from the galaxy centre and for different
kernel types (G: Gaussian, C: circular, A: \citetalias{2011PASP..123.1218A}, E: effective). Each column thus represents a vertical cut of the modeled
galaxy. Dashed lines indicate the characteristic scale height of
the intrinsic dust abundance profile. The top three panels show
scenarios \textit{a.}, \textit{b.} and \textit{c.} where the scale height varies and the SNR is kept constant at $10^{4}$. The two
bottom panels show scenarios \textit{b.1} and \textit{b.2} where the scale
height is fixed at 1 kpc and we vary the SNR
(see Table~\ref{table:scenarios} for a summary).
Each panel of Figure~\ref{fig:mosaic_deltaT} shows the
relationship between $\Delta \beta$ and $\Delta T$ just as in
Figure~\ref{fig:beta_temp}, and adds the spatial information to the data points.

Depending on the considered dust scale height, deviations from the reference
dust temperature and $\beta$ parameter reach $\pm 2$\,K and $\pm 0.6$,
respectively, with higher deviations for shorter scale heights and at higher
galactic latitudes. Regardless of the convolution kernel adopted, the negative
correlation is clearly measured for all the pixels in the vertical cut.

Scenarios \textit{a.}, \textit{b.} and \textit{c.} show that using the effective
kernels, very low ($< 1\%$) deviations from the reference values are obtained,
while the use of other convolution kernels lead to larger errors for $z \gtrsim
z_{\rm d}$. However, as expected, decreasing the level of SNR (scenarios
\textit{b.1} and \textit{b.2}), the noise starts to dominate over the signal and
very large discrepancies are observed in temperature and $\beta$, regardless of
the adopted convolution kernel.

\section{JWST PSF simulations}
\label{sec:jwst}

To show the reliability of our method for complex-shaped PSFs, we now
test our algorithm on simulated PSFs from a telescope with an uncommon optical
design. For this purpose, we select the Mid-InfraRed Instrument Imager (MIRI) of
the James Webb Space Telescope \citep[\textit{JWST},][]{bouchet2015mid}. The
\textit{JWST} main mirror being made of several hexagonal mirrors, the optical
response of this instrument shows highly non symmetrical features that need to
be accounted for in the homogenisation process.

\begin{figure}[t]
    \centering
    \includegraphics[width=0.45\textwidth]{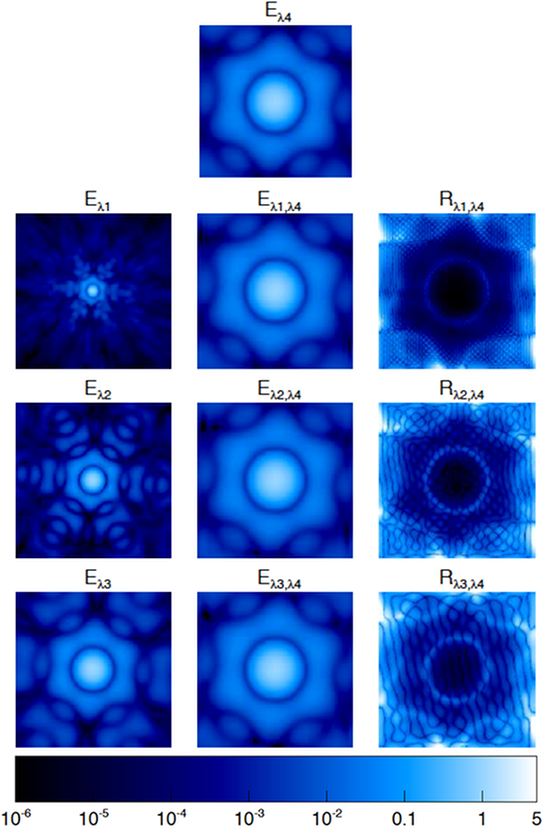}
    \caption{Proof of concept of PSF homogenisation for the \textit{JWST}/MIRI
    instrument. The PSFs from the first column, respectively at 5.6 $\mu$m, 11.3
    $\mu$m and 18.8 $\mu$m, are homogenised to the PSF on top of the second
    column at 25.5 $\mu$m, using \texttt{pypher} kernels. The resulting
    homogenised PSFs are shown in the second column. They are visually
    indistinguishable from the effective one on top. The relative residuals
    on the right confirm that the central part of the PSF is reconstructed at
    least to 0.1\%.}
    \label{fig:jwstkernels}
\end{figure}

We use WebbPSF \citep{perrin2012simulating}, the official \textit{JWST} PSF
simulation tool to generate a set of four broadband PSFs of $5''\times5''$
centered at $\lambda_{1}=5.6~\mu m$, $\lambda_{2}=11.3~\mu m$,
$\lambda_{3}=18.0~\mu m$ and $\lambda_{4}=25.5~\mu m$ in order to cover the
spectral range of MIRI. These broadband PSFs were generated from a set of 20
monochromatic ones assuming flat spectral energy distribution of the source, and
oversampled at 4 times the pixel size of the detector, corresponding to a pixel
scale of 0.11 arcsec. They are displayed on the first column (and top of the
second column) of Figure~\ref{fig:jwstkernels}.

Using \texttt{pypher}, we then compute three matching kernels, namely,
$K_{\lambda_1,\,\lambda_4},K_{\lambda_2,\,\lambda_4}$ and
$K_{\lambda_3,\,\lambda_4}$ (following the notation from
Section~\ref{sec:Herschel_PSFs}), to homogenise the first three PSFs to the
angular resolution of $E_{\lambda_4}$. Using the same procedure as in
Section~\ref{sub:kern_comparison}, we compare these homogenised PSFs to the
original one using the residual formalism \eqref{eq:residuals} applied to the
MIRI bandpasses. The resulting residual images $R_{\lambda_1,\,\lambda_4},
R_{\lambda_2,\,\lambda_4}$ and $R_{\lambda_3,\,\lambda_4}$ are shown on the
last column of Figure~\ref{fig:jwstkernels}.

The central region of the residual images, within the two main lobes of the
$E_{\lambda_4}$ PSF, has a low level of residuals ($\sim 10^{-5}$) in the
three configurations
Along the
image borders, there are some non-negligible residual patches. A quick visual
comparison with the homogenised PSFs (central column) shows these patches
correspond to extremely faint regions of the PSF ($< 10^{-6}$ w.r.t. the
peak) and thus having a very low impact in the matching process.


\section{Conclusions}
\label{sec:conclusion}

In this paper, we propose a new method for the generation of static PSF
homogenisation kernels which is applicable for instruments presenting complex
PSFs such as recent or future space-born telescopes.
\change{The PSF on such optical systems is hardly ever static over the field-of-view, but we restricted the purpose of this paper to the production of homogenization kernels for the study of regions of interest on the image, where the PSF can be considered non-variable. The treatment of the PSF varying over the whole field-of-view of modern instruments cannot be linearized as in this work and requires a very different approach. It will be the subject of a following paper.}
The application on
\textit{Herschel}/PACS and SPIRE and \textit{JWST}/MIRI instruments demonstrates
the performance of the proposed algorithm in terms of low residuals (better than
$10^{-2}-10^{-3}$ and $10^{-5}-10^{-6}$ for observed and simulated PSFs,
respectively).

To assess the improvement brought by our algorithm for multi-wavelength studies,
we address the estimation of dust temperature and spectral index $\beta$ of
astronomical objects using multi-band images taken in the submillimeter spectral
range by \textit{Herschel}. This estimation is made via pixel-by-pixel
measurements across these images which have different intrinsic angular
resolutions. Homogenisation kernels are thus traditionally used to bring all the
images to the same angular resolution. Most of the analysis performed so far use
either Gaussian kernels, or the circularised kernels produced by
\citet{aniano2011}. However, effective PSFs of space imagers are anisotropic, so
these methods are not accurate enough therefore introducing systematic
anti-correlation on $\beta$ and temperature measurements with an amplitude which
can be larger than the statistical noise.  We have checked that using \texttt{pypher}
kernels, systematic errors are in any case negligible compared to statistical
noise.

Finally, we provide the \texttt{pypher} software
\citep{alexandre_boucaud_2016_61392} to compute homogenisation
kernels to be used for current and future instruments.


\begin{acknowledgements}
AB would like to thank Hacheme Ayasso for useful discussions.
We acknowledge the CNES (Centre National d'\'Etudes Spatiales)
for supporting this work as part of the Euclid SGS (Science Ground Segment)
within the Euclid Consortium. We acknowledge the Euclid Consortium, the European
Space Agency and agencies and institutes supporting the development of Euclid.
Part of this work has received funding from the European Union's
Seventh Framework Programme (FP7/2007-2013) for the DustPedia project
(grant agreement n$^{\circ}$ FP7-SPACE-606847).
\change{This research made use of Astropy, a community-developed core Python
package for Astronomy (Astropy Collaboration, 2013).}
\end{acknowledgements}


\bibliographystyle{aa}
\bibliography{bib_conv_kern}

\begin{appendix}

\section{The \texttt{pypher} code}
\label{sub:pypher}

A Python code \change{called \texttt{pypher} \citep{alexandre_boucaud_2016_61392},} that
computes the static PSF homogenisation kernels described in
this work has been made publicly available and can be retrieved at \\
\url{https://github.com/aboucaud/pypher}.

Once installed, this program can be used through a command-line interface
taking as input the PSF images -- source and target -- as fits files, and
specifying the output filename for the kernel,
\begin{verbatim}
$ pypher psf_a.fits psf_b.fits kernel_a_to_b.fits
\end{verbatim}
The tunable parameters are
\begin{itemize}
    \item the regularisation parameter $\mu$ of the Wiener filter (see equation
    ~\ref{eq:wiener}) that penalises the high-frequencies, and should be set
    according to the image that will be homogenised,
    \item the input PSFs position angle with respect to their image to accurately
    take into account the shape of both PSF in the homogenisation process.
\end{itemize}

\noindent The program takes less than a second on a single CPU to compute a
kernel from two $512\times512$ PSF images.

\end{appendix}

\end{document}